\documentclass[conference]{IEEEtran}
\usepackage{ifpdf}

\usepackage{cite}

\ifCLASSINFOpdf
\else
\fi
\usepackage[cmex10]{amsmath}
\usepackage{array}

\usepackage{mdwmath}
\usepackage{mdwtab}

\usepackage{eqparbox}

\usepackage[tight,footnotesize]{subfigure}

\usepackage{fixltx2e}

\usepackage{stfloats}

\usepackage{url}

\begin{document}
%
\title{On the Cross-Correlation of a Ternary $m$-sequence of Period $3^{4k}-1$ and Its Decimated Sequence by $\frac{(3^{2k}+1)^{2}}{20}$}

\author{\IEEEauthorblockN{Yuhua Sun\IEEEauthorrefmark{1}\IEEEauthorrefmark{2}}
\and
\IEEEauthorblockN{Zilong Wang\IEEEauthorrefmark{1}}
\and
\IEEEauthorblockN{Hui Li\IEEEauthorrefmark{1}}
\and
\IEEEauthorblockA{
\IEEEauthorrefmark{1}Key Laboratory of Computer Networks and
 Information Security,
Xidian University,
 Xi'an 710071,
 Shaanxi, China \\
\IEEEauthorrefmark{2} College of Mathematics and
Computational Science,
China University of Petroleum,
Dongying 257061,
Shandong, China\\
}}


%


\maketitle

\begin{abstract}
Let $d=\frac{(3^{2k}+1)^{2}}{20}$, where $k$ is an odd integer. We show that the magnitude of the cross-correlation values of a ternary $m$-sequence $\{s_{t}\}$ of period $3^{4k}-1$ and its decimated sequence $\{s_{dt}\}$ is upper bounded by $5\sqrt{3^{n}}+1$, where $n=4k$. \\
\end{abstract}


%
\IEEEpeerreviewmaketitle

\section{INTRODUCTION}
One problem of considerable interest has been to find a decimation value $d$ such that the cross-correlation between a $p$-ary $m$-sequence $\{s_{t}\}$ of period $p^{n}-1$ and its decimation $\{s_{dt}\}$ is low. When $\mathrm{gcd}(d,p^{n}-1)=1$, the decimated sequence $\{s_{dt}\}$ is also a $m$-sequence of period $p^{n}-1$. Basic results on the cross-correlation between two $m$-sequences can be found in [1-3].

When $\mathrm{gcd}(d,p^{n}-1)\neq1$, the sequence $\{s_{dt}\}$ has period $\frac{p^{n}-1}{\mathrm{gcd}(d,p^{n}-1)}$. For this case, the reader is referred to [4-6]. Recently, Choi, No, and Chung \cite{7} investigated into the cross-correlation of a ternary $m$-sequence $\{s_{t}\}$ of period $3^{4k+2}-1$ and its decimated sequence $\{s_{dt}\}$ by $d=\frac{(3^{2k+1}+1)^{2}}{8}$. They derived the upper bound as $2\sqrt{3^{n}}+1$.

In this paper, we employ the method of \cite{7} to investigate the cross-correlation of a ternary $m$-sequence $\{s_{t}\}$ of period $3^{4k}-1$ and its decimated sequence $\{s_{dt}\}$ by $d=\frac{(3^{2k}+1)^{2}}{20}$, where $k$ is an odd integer. We show that the magnitude of the cross-correlation values is upper bounded by $5\sqrt{3^{n}}+1$, where $n=4k$.
\section{PRELIMINARIES}
Let $p$ be an odd prime, $\omega=e^{\frac{2\pi\sqrt{-1}}{p}}$ be a complex $p$th root of unity. Let $\{a_{t}\}$ and $\{b_{t}\}$ be two sequences of period $L$ over  $\mathrm{GF}(p)$. The cross-correlation between these two sequences at shift $\tau$ is defined by
$$C_{a,b}(\tau)=\sum\limits_{t=0}^{L-1}\omega^{a_{t+\tau}-b_{t}},$$
where $0\leq \tau < L$.

Let $\mathrm{GF}(p^{n})$ be the finite field with $p^{n}$ elements and $\mathrm{GF}(p^{n})^{*}=\mathrm{GF}(p^{n})\backslash\{0\}$.
The trace function $\mathrm{Tr}_{m}^{n}$ from the field $\mathrm{GF}(p^{n})$ onto the subfield $\mathrm{GF}(p^{m})$ is defined by
$$\mathrm{Tr}_{m}^{n}(x)=x+x^{p^{m}}+x^{p^{2m}}+\cdots+x^{p^{(h-1)m}},$$
where $h=n/m$.

Let $\alpha$ be a primitive element of $\mathrm{GF}(p^{n})$. A $p$-ary $m$-sequence $\{s_{t}\}$ is given by
$$s_{t}=\mathrm{Tr}_{1}^{n}(\alpha^{t}),$$
where $\mathrm{Tr}_{1}^{n}$ is the trace function from $\mathrm{GF}(p^{n})$ onto $\mathrm{GF}(p)$.
The periodic cross correlation function $C_{d}(\tau)$ between$ \{s_{t}\}$ and $\{s_{dt}\}$ is defined by
$$C_{d}(\tau)=\sum\limits_{t=0}^{p^{n}-2}\omega^{s_{t+\tau}-s_{dt}},$$
where $0\leq \tau \leq p^{n}-2$.

We will use the following notation unless otherwise specified. Let $\alpha$ be a primitive element of $\mathrm{GF}(3^{n})$, $n=4k$, $\omega=e^{\frac{2\pi\sqrt{-1}}{3}}$ and $d=\frac{(3^{2k}+1)^{2}}{20}$, where $k$ is an odd integer.
\section{CROSS-CORRELATION AND THE RANK OF THE QUADRATIC FORM}
In this section, we will give some results needed to prove our main result.
\newtheorem{theorem}{Theorem}
\newtheorem{lemma}{Lemma}
\newtheorem{corollary}{Corollary}
\newtheorem{remark}{Remark}
\newtheorem{example}{Example}
\begin{lemma}
Let the symbols be defined as above. Then the cross-correlation between $ \{s_{t}\}$ and $\{s_{dt}\}$ is given by
$$C_{d}(\tau)=-1+S_{d}(\tau).$$
Here,
$$2S_{d}(\tau)=\sum\limits_{x\in \mathrm{GF}(3^{n})}\omega^{q_{1}(x)}+\sum\limits_{x\in \mathrm{GF}(3^{n})}\omega^{q_{2}(x)},$$ where $$q_{1}(x)=\mathrm{tr}_{1}^{n}(ax^{3^{2(k+1)}+1}-x^{3^{2k}+1}),$$
$$q_{2}(x)=\mathrm{tr}_{1}^{n}(arx^{3^{2(k+1)}+1}-r^{d}x^{3^{2k}+1}),$$
 $a=\alpha^{\tau}$ and $r$ is a nonsquare in $\mathrm{GF}(3^{4})$.
\end{lemma}
\begin{IEEEproof}
By the definition of $C_{d}(\tau)$, we get
$$
\begin{array}{ll}
C_{d}(\tau)&=\sum\limits_{t=0}^{3^{n}-2}\omega^{s_{t+\tau}-s_{dt}}\\
&=\sum\limits_{t=0}^{3^{n}-2}\omega^{\mathrm{tr}_{1}^{n}(\alpha^{t+\tau}-\alpha^{dt})}\\
&=-1+\sum\limits_{x\in GF(3^{n})}\omega^{\mathrm{tr}_{1}^{n}(ax-x^{d})}\\
&=-1+S_{d}(\tau),
\end{array}
$$
where $a=\alpha^{\tau}$.

Since
$$\mathrm{gcd}(3^{2(k+1)}+1,3^{2k}+1)=2,$$
$$\mathrm{gcd}(3^{2(k+1)}+1,3^{2k}-1)=2,$$
and
$$3^{2(k+1)}+1\equiv 2\ \mathrm{mod}\ 4 ,$$ we have
$$\mathrm{gcd}(3^{2(k+1)}+1,3^{n}-1)=2.$$
It follows that $x^{3^{2(k+1)}+1}$ runs twice through all the squares in $\mathrm{GF}(3^{n})$ when $x$ runs through the nonzero elements of $\mathrm{GF}(3^{n})$. Let $r$ be a nonsquare in $\mathrm{GF}(3^{4})$. Then $r$ is also a nonsquare in $\mathrm{GF}(3^{n})$, since $\frac{n}{4}=k$ is odd. Therefore, $rx^{3^{2(k+1)}+1}$ runs twice through all the nonsquares in $\mathrm{GF}(3^{n})$ when $x$ runs through the nonzero elements of $\mathrm{GF}(3^{n})$. Furthermore, since $d(3^{2(k+1)}+1)=\frac{(3^{2(k+1)}-1)+20}{20}(3^{4k}-1)+3^{2k}+1\equiv 3^{2k}+1\ (\mathrm{mod}\ 3^{n}-1)$, we have
$$2S_{d}(\tau)=\sum\limits_{x\in \mathrm{GF}(3^{n})}\omega^{q_{1}(x)}+\sum\limits_{x\in \mathrm{GF}(3^{n})}\omega^{q_{2}(x)}.$$
\end{IEEEproof}
\begin{remark}
Let $\{\alpha_{1},\alpha_{2},\cdots,\alpha_{n}\}$ be a basis of $\mathrm{GF}(3^{n})$ over $\mathrm{GF}(3)$. Then $x$ can be expressed as $x=\sum\limits_{i=1}^{n}x_{i}\alpha_{i}$,
where $x_{i}\in\mathrm{GF}(3)$. It can be easily shown that $q_{1}(x)$ and $q_{2}(x)$ in Lemma 1 are of quadratic forms taking values in $\mathrm{GF}(3)$. For example,
$$
\begin{array}{ll}
q_{1}(x)&=\mathrm{tr}_{1}^{n}(ax^{3^{2(k+1)}+1}-x^{3^{2k}+1})\\
&=\mathrm{tr}_{1}^{n}(a\sum\limits\limits_{i=1}^{n}x_{i}\alpha_{i}^{3^{2(k+1)}}\sum\limits_{j=1}^{n}x_{j}\alpha_{j}\\
&\  \ \ -\sum\limits_{i=1}^{n}x_{i}\alpha_{i}^{3^{2k}}\sum\limits_{j=1}^{n}x_{j}\alpha_{j})\\
&=\mathrm{tr}_{1}^{n}(a\sum\limits_{i=1}^{n}\sum\limits_{j=1}^{n}x_{i}x_{j}\alpha_{i}^{3^{2(k+1)}}\alpha_{j}\\
&\ \ \ -\sum\limits_{i=1}^{n}\sum\limits_{j=1}^{n}x_{i}x_{j}\alpha_{i}^{3^{2k}}\alpha_{j})\\
&=\sum\limits_{i=1}^{n}\sum\limits_{j=1}^{n}(x_{i}x_{j})\mathrm{tr}_{1}^{n}(a\alpha_{i}^{3^{2(k+1)}}\alpha_{j}-\alpha_{i}^{3^{2k}}\alpha_{j})\\
&=\sum\limits_{i=1}^{n}\sum\limits_{j=1}^{n}(x_{i}x_{j})a_{ij},
\end{array}
$$
where $a_{ij}=\mathrm{tr}_{1}^{n}(a\alpha_{i}^{3^{2(k+1)}}\alpha_{j}-\alpha_{i}^{3^{2k}}\alpha_{j})$. Similarly, it can be shown that $q_{2}(x)$ is also of a quadratic form. In the remaining part of this paper, we always denote the quadratic forms in Lemma 1 as $q_{1}(x)$ and $q_{2}(x)$ respectively.
\end{remark}

To find the values of the exponential sum $S_{d}(\tau)$ in Lemma 1, we should determine the number of solutions $x$ such that $q_{i}(x)=c$ for any $c\in\mathrm{GF}(3)$, where $i=1,2$. Further, the number of solutions of $q_{i}(x)=c$ can be determined by the rank of the quadratic form $q_{i}(x)$, i.e., the number of variables of the form depends on. An alternative method to calculate the rank of a quadratic form is stated in the following lemma by M\"{u}ller \cite{4}.
\begin{lemma}\ \cite{4}
Let $f(x)\in \mathrm{GF}(p^n)[x]$ can be presented as a quadratic form in $\mathrm{GF}(p)[x_1, x_2, \cdots x_n]$. Furthermore, let
$$Y=\{y\in\mathrm{GF}(p^{n}):f(x+y)=f(x)\  \mathrm{for}\  \mathrm{all}\  x\in \mathrm{GF}(p^{n})\}.$$
Then rank($f$)=$n-\dim(Y)$.
\end{lemma}

From Lemma 2, we know that we should find the number of solutions $y\in\mathrm{GF}(3^{n})$ such that $q_{i}(x+y)=q_{i}(x)$ for all $x\in \mathrm{GF}(3^{n})$ to determine the rank of $q_{i}(x)$, where $i=1,2$. To this end, we have the following lemmas.
\begin{lemma}
The number of solutions $y\in \mathrm{GF}(3^{n})$ such that $q_{1}(x+y)=q_{1}(x)$ for all $x\in \mathrm{GF}(3^{n})$
equals the number of solutions $y\in \mathrm{GF}(3^{n})$ of
\begin{eqnarray}
a^{3^{2(k+1)}}y^{81}+y^{9}+ay=0,\label{1}
\end{eqnarray}
and the number of solutions $y\in \mathrm{GF}(3^{n})$ such that $q_{2}(x+y)=q_{2}(x)$ for all $x\in \mathrm{GF}(3^{n})$
equals the number of solutions $y\in \mathrm{GF}(3^{n})$ of
\begin{eqnarray}
(ar)^{3^{2(k+1)}}y^{81}-(r^{d}+r^{9d})y^{9}+ary=0,\label{2}
\end{eqnarray}
where $r$ is a nonsquare in $\mathrm{GF}(3^{4})$.
\end{lemma}
\begin{IEEEproof}
By the definition of $q_{1}(x)$, we have $q_{1}(x+y)=q_{1}(x)$ if and only if
\begin{eqnarray}
\mathrm{tr}_{1}^{n}(a(x+y)^{3^{2(k+1)}+1}-(x+y)^{3^{2k}+1})\nonumber\\
=\mathrm{tr}_{1}^{n}(ax^{3^{2(k+1)}+1}-x^{3^{2k}+1}),\nonumber
\end{eqnarray}
which implies
\begin{eqnarray}
\mathrm{tr}_{1}^{n}(x^{3^{2(k+1)}}(a^{3^{2(k+1)}}y^{81}+y^{9}+ay)\nonumber\\
+(ay^{3^{2(k+1)}+1}-y^{3^{2k}+1}))=0.\label{3}
\end{eqnarray}
Eq. (\ref{3}) holds for all $x\in \mathrm{GF}(3^{n})$ if and only if
\begin{eqnarray}
a^{3^{2(k+1)}}y^{81}+y^{9}+ay=0\label{4},
\end{eqnarray}
and
\begin{eqnarray}
\mathrm{tr}_{1}^{n}(ay^{3^{2(k+1)}+1}-y^{3^{2k}+1})=0.\label{5}
\end{eqnarray}

Next, we will show that Eq. (\ref{5}) is a consequence of Eq. (\ref{4}). From Eq. (\ref{4}) we obtain
\begin{eqnarray}
-y^{9}=a^{3^{2(k+1)}}y^{81}+ay.\label{6}
\end{eqnarray}
Raising the $3^{i}$ power for Eq. (\ref{6}) gives
\begin{eqnarray}
-y^{3^{2+i}}=a^{3^{2(k+1)+i}}y^{3^{4+i}}+a^{3^{i}}y^{3^{i}}.\label{7}
\end{eqnarray}
By the definition and the properties of trace function, we have
\begin{eqnarray}
&-\mathrm{tr}_{1}^{n}(y^{3^{2k}+1})\nonumber\\
&=-\mathrm{tr}_{1}^{n}(({y^{3^{2}}})^{3^{2k}+1})\nonumber\\
&\ \ =-\sum\limits_{i=0}^{n-1}(y^{3^{2}})^{3^{2k+i}+3^{i}}\nonumber\\
&
 \ \ \ \ \
  \ \ \ \ =\sum\limits_{i=0}^{n-1}(-y^{3^{2+i}})\cdot y^{3^{2(k+1)+i}}.\label{8}
\end{eqnarray}
Using Eq. (\ref{7}), we have
$$
\begin{array}{ll}
&\ \ \ -\mathrm{tr}_{1}^{n}(y^{3^{2k}+1})\\
&=\sum\limits_{i=0}^{n-1}(a^{3^{2(k+1)+i}}y^{3^{4+i}}+a^{3^{i}}y^{3^{i}})y^{3^{2(k+1)+i}}\\
&=\mathrm{tr}_{1}^{n}((ay^{3^{2(k+1)}+1})^{3^{2(k+1)}})+\mathrm{tr}_{1}^{n}(ay^{3^{2(k+1)}+1})\\
&=2\mathrm{tr}_{1}^{n}(ay^{3^{2(k+1)}+1})\\
&=-\mathrm{tr}_{1}^{n}(ay^{3^{2(k+1)}+1}),
\end{array}
$$
i.e., Eq. (\ref{5}) holds.
Thus, we only need to determine the number of solutions of Eq. (\ref{1}) to find the number of solutions of Eq. (\ref{3}).

By the similar argument, we can get the number of solutions $y\in \mathrm{GF}(3^{n})$ such that $q_{2}(x+y)=q_{2}(x)$ for all $x\in \mathrm{GF}(3^{n})$
equals the number of solutions $y\in \mathrm{GF}(3^{n})$ of
\begin{eqnarray}
(ar)^{3^{2(k+1)}}y^{81}-(r^{9d}+r^{d\cdot3^{2(k+1)}})y^{9}+ary=0.\label{9}
\end{eqnarray}
Since $r$ is a nonsquare in $\mathrm{GF}(3^{4})$ and $4|2(k+1)$, we have $r^{d\cdot3^{2(k+1)}}=r^{d}$. Hence Eq. (\ref{9}) is equivalent to Eq. (\ref{2}).
\end{IEEEproof}

From Lemma 3, in order to calculate the rank of $q_{1}(x)$ and $q_{2}(x)$ we have to determine the number of solutions $y\in \mathrm{GF}(3^{n})$ of Eqs. (\ref{1}) and (\ref{2}), respectively. Note that Eqs. (\ref{1}) and (\ref{2}) are both linearized forms, the possible number of solutions in $\mathrm{GF}(3^{n})$ for both equations are 1, $9$, or $81$.
\begin{lemma}
The
Eq. (\ref{2})
\begin{eqnarray}
(ar)^{3^{2(k+1)}}y^{81}-(r^{d}+r^{9d})y^{9}+ary=0\nonumber
 \end{eqnarray}
has $y=0$ as its only solution in $\mathrm{GF}(3^{n})$, where $r$ is a nonsquare in $\mathrm{GF}(3^{4})^{\ast}$.
\end{lemma}

\begin{IEEEproof}
Note that $r$ is a nonsquare in $\mathrm{GF}(3^{4})$, and that both $k$ and $\frac{3^{2k}+1}{10}$ are odd, we have $$r^{3^{2(k+1)}}=r,$$
$$r^{\frac{3^{4}-1}{2}}=r^{40}=-1$$
and
$$r^{40\cdot(\frac{3^{2k}+1}{10})^{2}}=r^{8\cdot\frac{(3^{2k}+1)^{2}}{20}}=r^{8d}=-1.$$
Therefore,
$$(ar)^{3^{2(k+1)}}=ra^{3^{2(k+1)}}$$
and
$$r^{9d}+r^{d}=r^{8d}r^{d}+r^{d}=-r^{d}+r^{d}=0.$$
Further, since $ra\neq0$, we get Eq. (\ref{2}) is equivalent to
$$y(a^{3^{2(k+1)}-1}y^{80}+1)=0.$$

Suppose $a^{3^{2(k+1)}-1}y^{80}=-1$. Note that $3^{2(k+1)}-1$ can be divided by $3^{4}-1=80$, we have
$$a^{3^{2(k+1)}-1}y^{80}=\beta^{80}=-1$$
for some $\beta\in\mathrm{GF}(3^{n})$, which implies that $80$ divides some odd multiples of $\frac{3^{4k}-1}{2}$. This is a contradiction because $160$ does not divide any odd multiples of $3^{4k}-1$.
\end{IEEEproof}

From the above lemmas, we know that $q_{1}(x)$ has the rank of $n$, $n-2$, $n-4$ and $q_{2}(x)$ has the rank of $n$.

\section{upper bound on cross-correlation magnitudes}
In this section, we will give the upper bound on the magnitude of the cross-correlation function $C_{d}(\tau)$ of the ternary $m$-sequence $ \{s_{t}\}$ and
 its decimated sequence $\{s_{dt}\}$ in Lemma 1. To this end, we will use the following definition and lemmas.

The quadratic character of $\mathrm{GF}(p^{n})$ is defined as
$$ \eta(x)=\left\{
\begin{array}{lll}
\ 1,\ \ \mathrm{if}\  x \mathrm{\ is\  a \ nonzero\  square\  in}\  \mathrm{GF}(p^{n})\\
-1,\ \ \mathrm{if}\  x \ \mathrm{is\  a\  nonsquare\  in}\  \mathrm{GF}(p^{n})\\
\ 0,\ \ \ \mathrm{if}\  x=0.
\end{array}
\right. $$
\begin{lemma}\ \cite{7}
 Let $\eta$ be the quadratic character of $\mathrm{GF}(p)$. Suppose that $f(x)$ is a nondegenerate quadratic form in $t$ variables with determinant $\triangle$.  Then the number of solutions $N(c)$ of $f(x)=c$ is given as follows:\\
 {\bf Case 1)} $t$ even;
 $$ N(c)=\left\{
\begin{array}{ll}
p^{t-1}-\epsilon p^{\frac{t-2}{2}},\ \ \ \ \ \ \ \mathrm{if}\  c\neq0\\
p^{t-1}+\epsilon (p-1)p^{\frac{t-2}{2}},\ \ \mathrm{if}\  c=0 \\
\end{array}
\right. $$
where $\epsilon=\eta((-1)^{\frac{t-1}{2}}\triangle)$.\\
{\bf Case 2)} $t$ odd;
 $$ N(c)=\left\{
\begin{array}{ll}
p^{t-1}-\epsilon p^{\frac{t-2}{2}},\ \ \ \ \ \ \ \mathrm{if}\  c\neq0\\
p^{t-1}+\epsilon (p-1)p^{\frac{t-2}{2}},\ \ \mathrm{if}\  c=0 \\
\end{array}
\right. $$
where $\epsilon=\eta((-1)^{\frac{t-1}{2}}\triangle)$.\\
\end{lemma}
\begin{lemma}\ \cite{7}
Let $\eta$ be the quadratic character of $\mathrm{GF}(3)$(i.e., $\eta(0)=0,$ $\eta(1)=1,$ $\eta(2)=-1$). Let $f(x)$ be a nondegenerate quadratic form in $t$ variables with determinant $\triangle$. Then
$$S=\sum\limits_{x\in GF(3^{n})}\omega^{f(x)}$$
is given by
$$ S=\left\{
\begin{array}{ll}
\epsilon3^{\frac{t}{2}},\ \ \ \mathrm{if}\  t\ \mathrm{is}\  \mathrm{even}\\
\epsilon i3^{\frac{t}{2}},\ \ \mathrm{if}\  t\ \mathrm{is}\  \mathrm{odd} \\
\end{array}
\right. $$
where $\epsilon=\eta((-1)^{\frac{t}{2}\triangle})$ for even $t$, $\epsilon=\eta((-1)^{\frac{t-1}{2}\triangle})$ for odd $t$.
\end{lemma}

Using Lemma 6, we can derive the upper bound of the magnitude of $C_{d}(\tau)$ in Lemma 1.

\begin{theorem}
The magnitude of the cross-correlation function $C_{d}(\tau)$ in Lemma 1 is upper bounded by
$$|C_{d}(\tau)|\leq5\cdot 3^{\frac{n}{2}}+1.$$
\end{theorem}
\begin{IEEEproof}
From Lemma 1, we know that
$$C_{d}(\tau)=-1+S_{d}(\tau)$$
and
$$2S_{d}(\tau)=\sum\limits_{x\in \mathrm{GF}(3^{n})}\omega^{q_{1}(x)}+\sum\limits_{x\in \mathrm{GF}(3^{n})}\omega^{q_{2}(x)},$$ where
$$q_{1}(x)=\mathrm{tr}_{1}^{n}(ax^{3^{2(k+1)}+1}-x^{3^{2k}+1}), $$ $$q_{2}(x)=\mathrm{tr}_{1}^{n}(arx^{3^{2(k+1)}+1}-r^{d}x^{3^{2k}+1}),$$
$a=\alpha^{\tau}$ and $r$ is a nonsquare in $\mathrm{GF}(3^{4})$. Let $\epsilon_{1}$ and $\epsilon_{2}$ denote the values given in Lemma 6 corresponding to $q_{1}(x)$ and $q_{2}(x)$, respectively. From \cite{8} we know that when the rank $t$ of a quadratic form is less than $n$, the corresponding exponential sum should be multiplied by $3^{n-t}$.

From Lemma 3, the possible rank combinations of $q_{1}(x)$ and $q_{2}(x)$ are $(n,n)$, $(n-2,n)$, $(n-4,n)$. Therefore, we should consider the following three cases to find the values of $S_{d}(\tau)$.\\
\\
{\bf Case 1)} Rank($q_{1})=n$ and rank($q_{2})=n$;\\
Using Lemma 6, we get
\begin{eqnarray}
2S_{d}(\tau)=\sum\limits_{x\in \mathrm{GF}(3^{n})}\omega^{q_{1}(x)}+\sum\limits_{x\in \mathrm{GF}(3^{n})}\omega^{q_{2}(x)}\nonumber\\
=(\epsilon_{1}+\epsilon_{2})3^{\frac{n}{2}}.
\end{eqnarray}

Hence, we have $|C_{d}(\tau)|=|-1+S_{d}(\tau)|\leq3^{\frac{n}{2}}+1.$\\
\\
{\bf Case 2)} Rank($q_{1})=n-2$ and rank($q_{2})=n$;\\
Using Lemmas 5 and 6, we have
\begin{eqnarray}
2S_{d}(\tau)=\sum\limits_{x\in \mathrm{GF}(3^{n})}\omega^{q_{1}(x)}+\sum\limits_{x\in \mathrm{GF}(3^{n})}\omega^{q_{2}(x)}\nonumber\\
=(3\epsilon_{1}+\epsilon_{2})3^{\frac{n}{2}}.
\end{eqnarray}

Therefore, we obtain $|C_{d}(\tau)|=|-1+S_{d}(\tau)|\leq2\cdot3^{\frac{n}{2}}+1.$\\
\\
{\bf Case 3)} Rank($q_{1})=n-4$ and rank($q_{2})=n$;\\
Using Lemmas 5 and 6, we get
\begin{eqnarray}
2S_{d}(\tau)=\sum\limits_{x\in \mathrm{GF}(3^{n})}\omega^{q_{1}(x)}+\sum\limits_{x\in \mathrm{GF}(3^{n})}\omega^{q_{2}(x)}\nonumber\\
=(9\epsilon_{1}+\epsilon_{2})3^{\frac{n}{2}}.
\end{eqnarray}

Hence, we obtain $|C_{d}(\tau)|=|-1+S_{d}(\tau)|\leq5\cdot3^{\frac{n}{2}}+1.$

As a conclusion, we have $|C_{d}(\tau)\leq5\sqrt{3^{n}}+1$.
\end{IEEEproof}






%


\end{document}